# Ultrahigh mobility electron gas formed at the BaO/SrTiO$_3$ interfaces

Cheng Cao[1,2+], Shengru Chen[1,2+], Jun Deng[1,2], Gang Li[1,3], Qinghua Zhang[1], Lin Gu[1], Tian-ping Ying[1,3], Er-Jia Guo[1,2,3], Jian-Gang Guo[1,3], Xiaolong Chen[1,2,3]

[1] Beijing National Laboratory for Condensed Matter Physics, Institute of Physics, Chinese Academy of Sciences, Beijing 100190, China.

[2] School of Physical Sciences, University of Chinese Academy of Sciences, Beijing 101408, China.

[3] Songshan Lake Materials Laboratory, Dongguan, Guangdong 523808, China.

[+] These authors contribute equally to the work.

*E-mail: ejguo@iphy.ac.cn; jgguo@iphy.ac.cn; chenx29@iphy.ac.cn

**Abstract**

Two-dimensional electron gas (2DEG) formed at the interface between two insulating oxides offers an opportunity for fundamental research and device applications. Binary alkaline earth metal oxides possess compatible lattice constants with both silicon and perovskite oxides, exhibiting an enormous potential to bridging those two materials classes for multifunctionalities. Here we report the formation of 2DEG at the interface between the rock-salt BaO and SrTiO$_3$. The highest electron mobility reaches 69000 cm$^2$ (V·s)$^{-1}$ at 2 K, leading to the typical Shubnikov–de Haas (SdH) oscillations under the high magnetic fields. The presence of SdH oscillations at different field-angles reveals a quasi-two-dimensional character of the Fermi surface. The first-principles calculations suggest that the effective charge transfer from the BaO to Ti 3$d_{xy}$ orbital at the interfaces is responsible to the observed high carrier mobility. Our results demonstrate that the BaO/STO heterointerface is a platform for exploring the correlated quantum phases, opening a door to the low-power and mesoscopic electronic devices.

**Introduction**

Two-dimensional electron gas (2DEG) confined at the interface between two insulating oxides exhibits the exotic properties such as high carrier mobility[1], interfacial magnetism[2,3], superconductivity[4,5], and *etc*. Extensive effort, including combining versatile materials[4,5], optimizing polarization of termination[6,7], tuning lattice mismatch[8], and controlling oxygen vacancies[9], has been spent to enhance carrier mobility and to understand the origin of above-mentioned emergent phenomena. In the most intensively-studied $LaAlO_3$(LAO)/$SrTiO_3$(STO) interfaces[10,11], the polar planes in LAO and non-polar planes in STO match together, resulting in a large compressive strain of 2.97%[1]. Astonishingly, the LAO/STO interfaces exhibit an unexpected high mobility of 1000−6600 $cm^2 (V·s)^{-1}$ at 2 K[12]. It is generally ascribed to the polar discontinuity at the interfaces that leads to the electronic reconstruction between the electron charged $(LaO)^+$ layer and neutral $(TiO_2)^0$ layer[13]. The internal electric potential exceeds the $E_g$ of STO and then drives the electrons from the LAO valence band into the STO conduction band. The charged defects at the interfaces are one of the factors that play a key role in limiting the charge mobility. Over decades, lots of attempts have been taken to minimize the defect formation at the 2DEG interfaces. Huang *et al.* had replaced the conventional LAO layer with a lattice-compatible $(La_{0.3}Sr_{0.7})(Al_{0.65}Ta_{0.35})O_3$ (LSAT) layer. The lattice mismatch between STO and LSAT reduces to less than 1%, leading to a significantly enhanced carrier mobility of 35000 $cm^2 (V·s)^{-1}$ at 2 K[8]. Similarly, a record high carrier mobility up to 140000 $cm^2 (V·s)^{-1}$ was achieved at the interfaces between the spinel-type γ-$Al_2O_3$ and STO, in which a small lattice mismatch of 1.2% manifests the well-defined interfaces[14].

To build the high-quality interfaces with the minimal misfit strain is highly desirable for improving the carrier mobility in 2DEGs. In terms of the crystallography, the STO substrate can be viewed as a stack of SrO and $TiO_2$ layers, where the upmost surface-plane terminates with a $TiO_2$ layer. The intuition option is to epitaxially grow AO (A = Ca, Sr, and Ba) film on the $TiO_2$-layers directly[15]. Therefore, the binary alkaline earth metal oxides possess the rock-salt crystal structure with lattice parameters between 4.2 and 5.6 Å. Among them, the lattice parameter of BaO (5.539 Å)[16] is close to the √2*$a_{STO}$ (5.523 Å), yielding to a compressive strain as small as −0.3% along the diagonal direction. Therefore, BaO is an ideal candidate for exploring its intrinsic physical properties and investigating the emergent phenomena at the interfaces[17, 18]. Furthermore, BaO has a compatible lattice constant with Si ($a$ = 5.43 Å), thus it can be served as a buffer layer to bridging the functional oxides with matured CMOS techniques.

Here we report the epitaxially growth of BaO thin films on STO substrates. The carrier mobility at the BaO/STO interfaces reaches 69000 cm$^2$ (V·s)$^{−1}$ and the sheet carrier concentration is ~ 10$^{16}$ cm$^{-2}$ at 2 K. The typical Shubnikov–de Haas (SdH) oscillations were observed under the high magnetic fields. The itinerate carriers located at the BaO/STO interfaces are responsible for the observed exotic transport behaviors, further confirmed by our theoretical calculations.

**Results**

**Sample growth and structural characterizations.**

The BaO films were grown on the STO substrates by pulsed laser deposition (see Methods). After the growth, a 10-nm-thick STO layer was capped on top of the BaO films to prevent the undesired deterioration of BaO films from the moistures in air. X-ray diffraction (XRD) were performed to check the crystallinity of the BaO films. Figure 1b shows a typical XRD $θ$-$2θ$ curve of a BaO thin film grown on a STO substrate. The out-of-plane lattice constant of the epitaxial BaO films is 5.528 Å. Only 00$l$ reflections from the BaO and STO are observed, indicating that the film is epitaxially grown on the substrate (Fig. 1b). The rocking curve of BaO 002 peak has a full width at half maximum (FWHM) value of 0.338º (inset of Fig. 1b), confirming an excellent crystalline quality of the as-grown films. Figure 1c shows the in plane $φ$-scans of the BaO film and STO substrate around the (113) plane. Four sharp and discrete peaks equally separate by 90º, revealing that both film and substrate have a fourfold rotational symmetry, in agreement with their cubic structures. Note that the BaO 113 peak shifts by 45º compared to that of STO, providing the solid evidence for rotating BaO unit cell in-plane toward the diagonal of the STO, i. e. BaO [110]//STO [100] (Fig. 1a), with minimized misfit strain. Furthermore, we performed the cross-sectional high-angle annular dark-filed (HAADF) imaging on the BaO films using scanning transmission electron microscopy (STEM). Since the BaO is extremely sensitive to the moisture in air during the specimen preparation, only a few unit cells of BaO at the interfaces are captured. In the HAADF-STEM images, the intensity scales with the value of $Z^{1.7}$, where $Z$ is the atomic number of elements. The brighter spots correspond to the heavier elements. In Fig. 1d, the brighter spots represent the Ba atoms, and the weaker spots are the Sr atoms, followed by the Ti atoms. The STEM measurements reveal that the high degree of coherent structure maintaines at the interfaces between BaO and STO, providing the solid basis for BaO thin film growth. Figure 1e shows the annular bright field (ABF)-STEM image at the BaO/STO interface. The high-magnified ABF image helps to identify the positions of the O atoms and further confirm the orientation and epitaxial growth of BaO thin films.

**Electrical transport properties**

To investigate the transport behaviors of BaO films, we firstly measured the temperature-dependent sheet resistance ($R_{xx}$) of the BaO films under zero magnetic field (Fig. 2a). Strikingly, the BaO films exhibit a superior metallic behavior down to 2 K with ultralow residue resistance of $7.3 \times 10^{-4}$ Ω/□, which is comparable to that of bulk Cu[19] and Sr$_2$CrWO$_6$ (4.9–92 nΩ·cm at 2 K)[20]. The residue resistance rations [$RRR = R_{xx}(300\ \text{K})/R_{xx}(2\ \text{K})$] of the BaO films are 140-4200. An ultrahigh $RRR$ is a rare case, indicating that the BaO films possess a superior crystalline quality and are ideal highly pure metallic system. To exclude the effect of oxygen vacancies to the transport results, we compared the $R_{xx}$-$T$ curves of the BaO films deposited at a higher growth temperature of 750 °C. Typically, the higher growth temperature will make the system get more activated energy, leading to the increased oxygen vacancies in both films and substrates. We find that the BaO films transit from a metallic phase to a narrow band-gap semiconductor (Supplementary Figs. S1). These results strongly suggest that the observed metallicity is intrinsic to BaO, ruling out the possible contribution from oxygen vacancies in STO substrates[21, 22, 23].

Having established the intrinsic transport behavior of BaO films, we further conducted the Hall measurements to understand its electronic characteristics. For the magnetic fields ($\mu_0 H$) below 5 T, the Hall resistance ($R_{xy}$) exhibits the linear-like behaviors at all temperatures, as shown in Fig. 2b. The carrier density (*n*) and mobility (*μ*) are extracted by fitting the $R_{xy}$-*H* curves using a single-band model. Figs. 2c and 2d show the *n* and *μ* as a function of applied fields, respectively. With decreasing temperature from 300 to 2 K, *n* reduces from $1.1 \times 10^{17}$ cm$^{-2}$ to $2.7 \times 10^{16}$ cm$^{-2}$. Please note that *n* of BaO films is at least an order of magnitude higher than those of the other oxide films grown on STO substrates[14], suggesting that the substrate does not dominate the free carriers in the system. The calculated *μ* at 300 K is 121 cm$^2$ (V·s)$^{-1}$, which is higher than the typical *μ* [2-12 cm$^2$ (V·s)$^{-1}$] in the most STO-based 2DEGs. At 2 K, the *μ* reaches up to 69000 cm$^2$(V·s)$^{-1}$. The observed ultrahigh carrier mobility in the BaO films is comparable to those reported in bulk Sr$_2$Cr*M*O$_6$ (*M* = Mo, W)[20, 24], graphene embedded in metals[25], and Ca$_2$N[26]. In addition, the magnetoresistance (MR), shows linearly increasing with the magnetic field, with no sign of saturation, and reached more than 350% at ±5 T and 2 K. The linear magnetoresistance (LMR) may be consistent with the long electron mean free path and high mobility. With increasing temperature, the value of MR decreases dramatically as shown in Supplementary Fig. S2.

**Observation of quantum oscillations in the BaO films**

The coexistence of both high *n* and *μ* in the BaO films is a rare case. Therefore, we further explore the intrinsic states of BaO films under the magnetic fields. Principally, high *RRR* values and high carrier mobility have enabled observations of quantum oscillations via the electrical resistivity measurements. Firstly, we measured the Hall resistance ($R_{xy}$) (Fig. 3a) and magnetoresistance [($R_H$-$R_0$)/$R_0$] (Fig. 3d) at *T* = 0.3 – 20 K up to 18 T. The magnetic fields were applied perpendicular to the surface plane (*H // c*). The BaO films show typical Shubnikov–de Haas (SdH) oscillations in both $R_{xy}$ and ($R_H$-$R_0$)/$R_0$ when *T* below 3 K and $\mu_0 H$ above 7 T. The quantum oscillations become significant by plotting their amplitude values ($\Delta R_{xy}$) against $B^{-1}$ after subtracting a polynomial background contributed from the normal and high-order Hall effects (Fig. 3b). Figure 3c shows the Fourier transform (FT) analysis of the $\Delta R_{xy} - B^{-1}$ curves measured at 0.3 – 2.7 K. Two main components of the oscillations with frequency $F_\alpha$ = 51 T and $F_\beta$ = 78 T are deducted. And the amplitudes of these two components decay gradually to zero with increasing temperature to 3 K. The nonlinear background of Hall resistance can be explained with a two-carrier model by introducing an additional electron. We calculate the effective mass of two types of electrons using the Lifshitz-Kosevich (*L-K*) equation,

$$RT = A'^{\left(\frac{m^*}{m_0}\right)}T \Big/ sinh\left[A'^{\left(\frac{m^*}{m_0}\right)}T\right]$$

where $A' = 2\pi^2 k_B m^*/e\hbar H_{eff}$ and $H_{eff} = 2/(1/H_1 + 1/H_2)$, in which $H_1$ and $H_2$ are the lower and upper limits of the magnetic field range of oscillations, respectively. $R_T$, $m_0$, $m^*$, $k_B$ and $\hbar$ are the amplitude of the fast Fourier transformation (FFT), free electron mass, effective mass, Boltzmann constant, and Planck constant, respectively. By fitting the temperature-dependent oscillation amplitude (inset of Fig. 3c), we obtained the effective masses of two types of electrons are $m_\alpha^* = (1.15 \pm 0.12)m_0$ and $m_\beta^* = (1.31 \pm 0.18)m_0$, respectively. These values are comparable to the effective masses of other STO-based 2DEGs [14, 23, 27, 28]. Simultaneously, the Dingle temperature ($T_D$) and the total scattering time ($\tau$) are determined to be 1.31 K (0.94 K) and 1.17 ps (1.29 ps) for $F_\alpha$ ($F_\beta$), respectively. The long scattering time provides independent evidence to the high mobility in the BaO films.

In addition, we rotated the direction of the applied magnetic fields with respect to the [001] axis, *i. e.* surface normal, and measured the quantum oscillations at various angles ($\theta$). Fig. 4a shows the *B*-dependent ($R_H$-$R_0$)/$R_0$ curves. The maximum value in the ($R_H$-$R_0$)/$R_0$ is 1300% at 18 T and 0.3 K as the *B* is parallel to the surface normal. The magnitude and period oscillations of ($R_H$-$R_0$)/$R_0$ gradually decreases and weakens as increasing the $\theta$. We notice that the quantum oscillations persist up to 59° and the frequency follows the 1/cos$\theta$ scaling as shown in Fig. 4b. This feature suggests that the Fermi surface of BaO film is quasi-two-dimensional type. When the *B* is parallel to the *ab*-plane, the oscillations of ($R_H$-$R_0$)/$R_0$ disappears, indicating the anisotropic Fermi surface of BaO film. In addition, the ($R_H$-$R_0$)/$R_0$ of BaO film exhibits the parabolic-like behavior at the lower fields and the linear dependence of at the higher field, see Fig. 3d and Fig. 4a like the observations in $Sr_2CrMoO_6$ [20, 24].

**Band structure of BaO/STO interface calculated by first-principles calculations**

The first-principles calculations were conducted to get a better understanding of the 2DEG on the interface between STO and BaO. We carefully optimize the interface structure by minimizing the formation energy (Supplementary Fig. S3) and compare the local density of states (LDOS) for each layer (Supplementary Fig. S4). As increasing the layer thickness, the lineshape of the LDOS does not change until the thickness of BaO and STO layers beyond 3 and 5 u.c., respectively. Therefore, we construct a supercell by stacking 3.5 u.c.-thick BaO over a 5.5 u.c-thick STO (Fig. 5a). The upper panel of Fig. 5(b) displays the corresponding density of states (DOS), demonstrating the BaO/STO interfaces keep the insulating state. To declare the oxygen vacancy effect on the 2DEG, we create one and two oxygen vacancies of BaO layer near the surface in a 2×2 superlattice (Supplementary Fig. S5). At this time, metallicity emerges as evidenced by the DOS plot in the bottom panel of Fig. 5(b). The layer projected DOS shows that the states across the Fermi level mainly come from the interface (Fig. S6), and these electrons are composed of Ti-$3d_{xy}$ orbitals (Fig. 5(c)). This was also confirmed by the electronic structure in Fig. S7. With these calculations, we can obtain an insight to the formation of 2DEG of STO/BaO interface. Firstly, the perfect interface is insulating while it is metallic with oxygen vacancies. So here oxygen vacancies play a key role in inducing free carriers. Then, these carriers will transfer to the Ti-$d_{xy}$ orbitals since under octahedron crystal field Ti-$t_{2g}$ orbital is empty in STO. Hence, the effect of oxygen vacancies resembles electron doping. And the electrons on the Ti-$t_{2g}$ orbitals are responsible for the observed quantum oscillations.

In summary, we show that a new type of oxide-2DEG formed at the BaO/STO interfaces. The unique ultrahigh mobility and carrier density are observed simultaneously, leading to the nonlinear magnetoresistance and distinct quantum oscillations under high magnetic fields and low temperatures. We believe that the possible charge transfer at the interfaces, confirmed by our first-principles calculations, would be responsible to the occurrence of observed quantum phenomena. This work suggests that the rock-salt structured *p*-electron oxides show a great promise for designing oxide-based electronics with superior properties and further stimulate the investigations of *p-d* electron coupling at the oxide interfaces.

**Methods**

**Sample preparation.** High-quality BaO thin films were epitaxially grown on (001)-oriented SrTiO$_3$ (STO) substrates by pulses laser deposition. The STO substrates were pretreated by buffered HF and annealed at a high-temperature of 1100 °C for 1.5 hours to ensure TiO$_2$-terminated surfaces. Before deposition, the base pressure was initially evacuated to 10$^{-8}$ Torr. A 20 nm-thick BaO layer was firstly deposited in vacuum at a substrate temperature of 650-750 °C using a polycrystalline BaO$_2$ pellet. The as-grown BaO film was *in-situ* annealed in vacuum at the growth temperature for 1 h. The annealing process was essential for crystallizing the BaO film. After annealing, a 10 nm-thick STO was deposited as a capping layer to prevent the BaO films from reacting with H$_2$O and CO$_2$ in air.

**Structural characterizations.** The lattice parameters and crystalline quality of BaO films were determined by X-ray diffraction (XRD) and Phi-scans. XRD measurements were conducted on a high-resolution four-circle diffractometer with Cu *K*α radiation (Bruker D8 Discovery). The Phi-scans were taken around the 113 reflections of the BaO films and STO. TEM specimens were prepared using the standard ion beam (FIB) lift-out process. HAADF and ABF imaging were performed in the STEM mode using JEM ARM 200CF microscopy at Institute of Physics (IOP) of Chinese Academy and Science (CAS).

**Transport measurements.** Transport properties of the BaO films were measured using a four-probe Van der Pauw method. The electrodes are ultrasonically wire-bonded Al wires and placed at the corners of the square samples. The BaO films can be contacted directly using wire-bonding through the STO capping layer. Temperature-dependent resistance and Hall measurements were conducted on a Physical Property Measurement System (PPMS) in the temperature range between 2 to 300 K. For the transport measurements with temperatures below 2 K were performed in a sorption pumped $^3$He cryostat with standard lock-in technique. The magnetic field of the $^3$He cryostat was applied up to 18 T. During all the transport measurements, the applied currents were 200 μA (for AC current, the frequency was 30.99 Hz) to avoid the heating effects.

**Computational calculation methods.** The first-principles calculations were carried out with the density functional theory (DFT) implemented in the Vienna *ab* initio simulation package (VASP) [33]. We adopted the generalized gradient approximation (GGA) in the form of the Perdew-Burke-Ernzerhof (PBE) [34] for the exchange-correlation potentials. The projector augmented-wave (PAW) [35] pseudopotentials were used with a plane wave energy 500 eV. A Monkhost-Pack [36] Brillouin zone sampling grid with a resolution $0.02 \times 2\pi$ Å$^{-1}$ was applied. The atomic positions and lattice parameters were relaxed until all the strain forces on each atom were less than $10^{-2}$ eV/Å, which is close to their bulk values. For the BaO/STO interface model without oxygen vacancies, the atomic positions were relaxed until all the forces on the atoms near the interface were less than $5 \times 10^{-2}$ eV/Å and a $9 \times 9 \times 1$ k-mesh was used in sampling the Brillouin. For BaO/STO interface model with oxygen vacancies ($2 \times 2 \times 1$ superlattice of BaO/STO). Only Γ point was used in the Brillouin sampling.

**References**

1. Ohtomo A, Hwang HY. A high-mobility electron gas at the LaAlO$_3$/SrTiO$_3$ heterointerface. *Nature* **427**, 423-426 (2004).

2. Kalisky B*, et al.* Critical thickness for ferromagnetism in LaAlO$_3$/SrTiO$_3$ heterostructures. *Nat Commun* **3**, 922, (2012).

3. Herranz G*, et al.* Full oxide heterostructure combining a high-T$_C$ diluted ferromagnet with a high-mobility conductor. *Phys Rev B* **73**, (2006).

4. Anh LD*, et al.* High-mobility 2D hole gas at a SrTiO$_3$ Interface. *Adv Mater* **32**, e1906003 (2020).

5. Lu D, Baek DJ, Hong SS, Kourkoutis LF, Hikita Y, Hwang HY. Synthesis of freestanding single-crystal perovskite films and heterostructures by etching of sacrificial water-soluble layers. *Nat Mater* **15**, 1255-1260 (2016).

6. Cantoni C*, et al.* Electron transfer and ionic displacements as the origin of the 2D electron gas at the LAO/STO Interface: Direct measurements with atomic-column spatial resolution. *Adv Mater* **24**, 3952-3957 (2012).

7. Bark CW, *et al.* Switchable induced polarization in LaAlO$_3$/SrTiO$_3$ heterostructures. *Nano Lett* **12**, 1765-1771 (2012).

8. Huang Z, *et al.* The effect of polar fluctuation and lattice mismatch on carrier mobility at oxide interfaces. *Nano Lett.* **16**, 2307-2313 (2016).

9. Kalabukhov A, Gunnarsson R, Börjesson J, Olsson E, Claeson T, Winkler D. Effect of oxygen vacancies in the SrTiO$_3$ substrate on the electrical properties of the LaAlO$_3$/SrTiO$_3$ interface. *Phys Rev B* **75**, (2007).

10. Noland JA. Opical absorption of single-crystal strontium titanate. *Phys Rev B* **94**, 724-724 (1954).

11. van Benthem K, Elsasser C, French RH. Bulk electronic structure of SrTiO$_3$: Experiment and theory. *J Appl Phys* **90**, 6156-6164 (2001).

12. Caviglia AD, *et al.* Two-dimensional quantum oscillations of the conductance at LaAlO$_3$/SrTiO$_3$ interfaces. *Phys Rev Lett* **105**, 236802 (2010).

13. Nakagawa N, Hwang HY, Muller DA. Why some interfaces cannot be sharp. *Nat Mater* **5**, 204-209 (2006).

14. Chen YZ, *et al.* A high-mobility two-dimensional electron gas at the spinel/perovskite interface of gamma-Al$_2$O$_3$/SrTiO$_3$. *Nat Commun* **4**, 1371 (2013).

15. Gagnidze T, Ma H, Cancellieri C, Bona GL, La Mattina F. Structural properties of ultrathin SrO film deposited on SrTiO$_3$. *Sci Technol Adv Mater* **20**, 456-463 (2019).

16. Hubbard KJ, Schlom DG. Thermodynamic stability of binary oxides in contact with silicon. *J Mater Res* **11**, 2757-2776 (1996).

17. Bousquet E, Spaldin NA, Ghosez P. Strain-induced ferroelectricity in simple rocksalt binary oxides. *Phys Rev Lett* **104**, 037601 (2010).

18. Nascimento VB, da Costa BV, Rino JP. Novel ferroelectric phase in bulk BaO obtained by application of anisotropic strain. *Appl Phys A Mater Sci Process* **126**, 744 (2020).

19. Matula RA. Electrical-resistivity of copper, gold, palladium, and siliver. *J Phys Chem Ref Data* **8**, 1147-1298 (1979).

20. Wang Z-C, *et al.* Giant linear magnetoresistance in half-metallic Sr$_2$CrMoO$_6$ thin films. *NPJ Quantum Mater* **6**, (2021).

21. Tufte ON, Chapman PW. Electron mobility in semiconducting strontium titanate. *Phys Rev* **155**, 796-802 (1967).

22. Szot K, Speier W, Carius R, Zastrow U, Beyer W. Localized metallic conductivity and self-healing during thermal reduction of $SrTiO_3$. *Phys Rev Lett* **88**, 075508 (2002).

23. Jalan B, Stemmer S, Mack S, Allen SJ. Two-dimensional electron gas in δ-doped $SrTiO_3$. *Phys Rev B* **82**, (2010).

24. Zhang J, *et al.* Giant positive magnetoresistance in half-metallic double-perovskite $Sr_2CrWO_6$ thin films. *Sci Adv* **3**, 7 (2017).

25. Cao M, *et al.* Ultrahigh electrical conductivity of graphene embedded in metals. *Adv Funct Mater* **29**, 17 (2019).

26. Lee K, Kim SW, Toda Y, Matsuishi S, Hosono H. Dicalcium nitride as a two-dimensional electrode with an anionic electron layer. *Nature* **494**, 336-340 (2013).

27. Rubi K, *et al.* Aperiodic quantum oscillations in the two-dimensional electron gas at the $LaAlO_3/SrTiO_3$ interface. *NPJ Quantum Mater* **5**, 9 (2020).

28. Kim M, Bell C, Kozuka Y, Kurita M, Hikita Y, Hwang HY. Fermi surface and superconductivity in low-density high-mobility delta-doped $SrTiO_3$. *Phys Rev Lett* **107**, 106801 (2011).

29. Jin KX, *et al.* Tunable photovoltaic effect and solar cell performance of self-doped perovskite $SrTiO_3$. *AIP Adv* **2**, 042131 (2012).

30. Siemons W, *et al.* Origin of charge density at $LaAlO_3$ on $SrTiO_3$ heterointerfaces: possibility of intrinsic doping. *Phys Rev Lett* **98**, 196802 (2007).

31. Liu ZQ, *et al.* Origin of the two-dimensional electron gas at $LaAlO_3/SrTiO_3$ Interfaces: The role of oxygen vacancies and electronic Reconstruction. *Phys Rev X* **3**, 021010 (2013).

32. Mayorova AF, Mudretsova SN, Mamontov MN, Levashov PA, Rusin AD. Thermoanalysis od the system $BaO_2$-$BaO$-$O_2$. *Thermochim Acta* **217**, 241-249 (1993).

33. Kresse G, Furthmuller J. Efficiency of ab-initio total energy calculations for metals and semiconductors using a plane-wave basis set. *Comput Mater Sci* **6**, 15-50 (1996).

34. Perdew JP, Burke K, Ernzerhof M. Generalized Gradient Approximation Made Simple. *Phys Rev Lett* **78**, 1396-1396 (1997).

35. Kresse G, Joubert D. From ultrasoft pseudopotentials to the projector augmented-wave method. *Phys Rev B* **59**, 1758-1775 (1999).

36. Monkhorst HJ, Pack JD. Special points for Brillouin-zone integrations. *Phys Rev B* **13**, 5188-5192 (1976).

**ACKNOWLEDGEMENTS**

This work is financially supported by the MoST-Strategic International Cooperation in Science, Technology and Innovation Key Program (Grant No. 2018YFE0202600), the National Key Research and Development Program of China (Grant Nos. 2016YFA0300600 and 2020YFA0309100), the Natural Science Foundation of China (Grant Nos. 51922105, 51532010, 11974390), the Beijing Natural Science Foundation (Grant Nos. Z200005, Z190010 and 2202060), the Strategic Priority Research Program of Chinese Academy of Sciences (Grant No. XDB33030200), and the Beijing Nova Program of Science and Technology (Grant No. Z191100001119112).

**AUTHOR CONTRIBUTIONS**

This work was initiated and supervised by X.L.C. and J.G.G. Samples were grown and processed by S.R.C., C.C., under the guidance of E.J.G.; The first-principles calculations were performed by J.D. and J.G.G.; TEM experiments were performed by Q.H.Z. and L.G.; Transport measurements at ultralow temperatures were performed by G.L. C.C., J.G.G. and E.J.G. wrote the manuscript with input from all coauthors.

**COMPETING INTERESTS**

The authors declare no competing interests.

**Figures and figure captions**

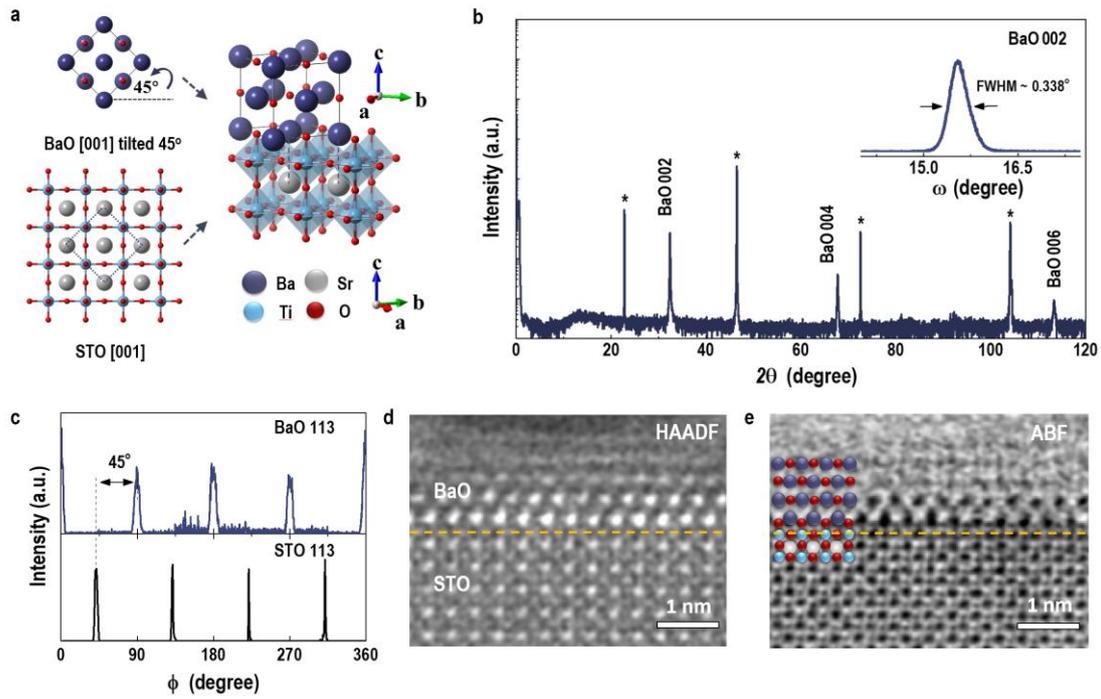

**Figure1. Crystal structure and epitaxial growth of BaO films**. (**a**) Schematic of the crystal structures and epitaxial growth of BaO films on the STO substrates. The unit cell of BaO rotates in-plane by 45º with respect to that of STO. (**b**) XRD $\theta$-$2\theta$ scan of a BaO film grown on a STO substrate. Insert shows the rocking curve around the BaO 002 reflection, yielding a FWHM ~ 0.338º. (**c**) Phi scans around the 113 reflections of a BaO film (top) and a STO substrate (bottom). (**d**) Cross-sectional HAADF image and (**e**) ABF image at the BaO/STO interface taken in the STEM mode. Inset shows the illustrate of the atomic arrangement across the interfaces.

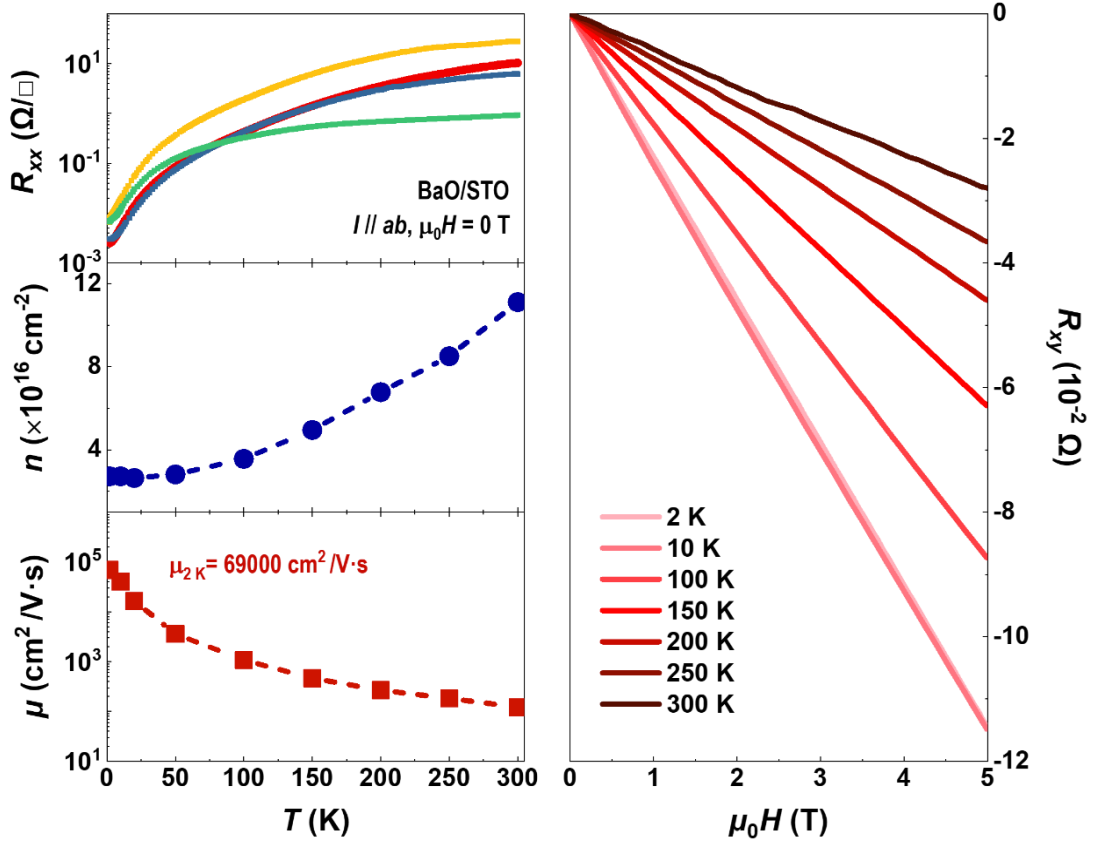

**Figure 2. Transport properties of the BaO films.** (**a**) Temperature-dependent sheet resistance ($R_{xx}$) of the BaO films with a large variation in *RRR* values up a maximum of 4240 is observed. (**b**) Field dependence of Hall resistance ($R_{xy}$) of BaO films at various temperatures. (**c**) Carrier density (*n*) and (**d**) mobility (*μ*), derived from Hall measurements using a single band model, plot as a function of temperature.

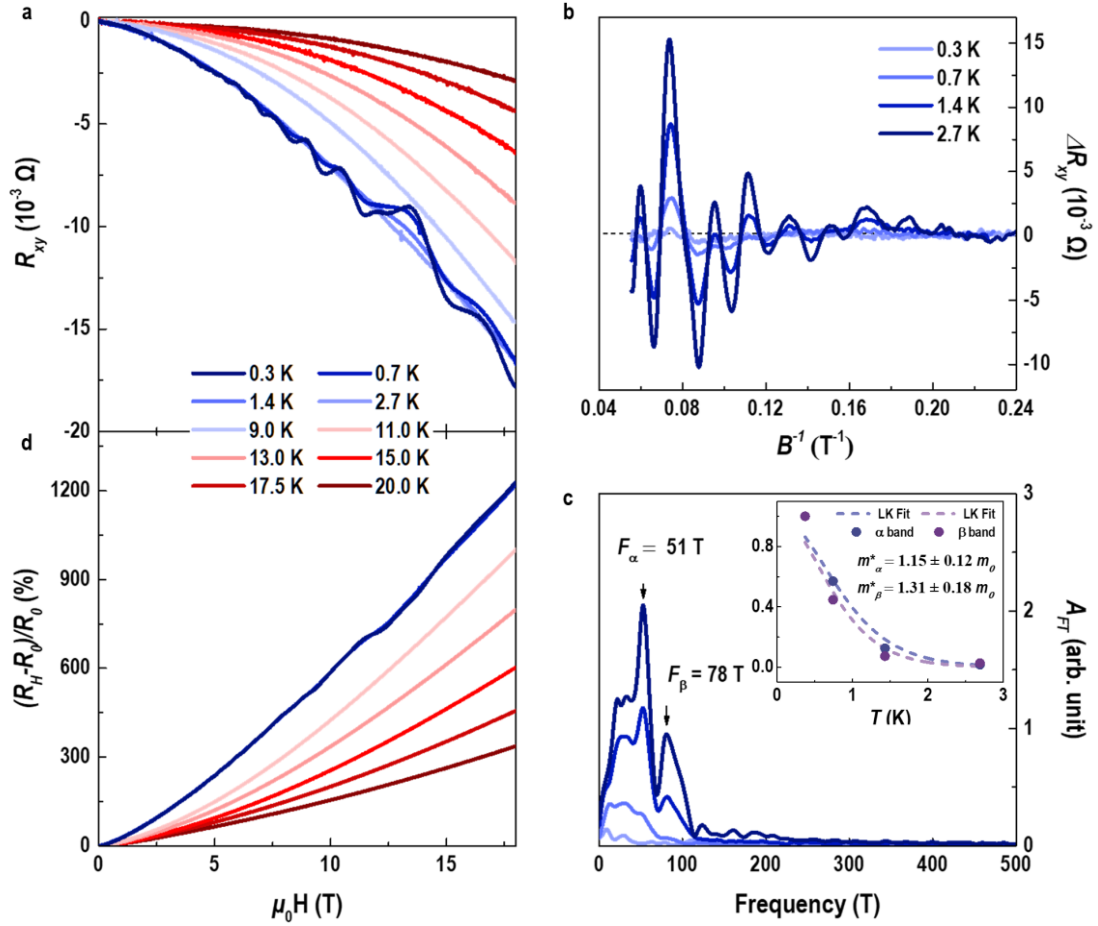

**Figure 3. Observation of quantum oscillation in the BaO films.** (**a**) $R_{xy}$ as a function of magnetic fields at various temperatures. (**b**) Shubnikov–de Haas (SdH) oscillations at various temperatures. (**c**) Fast Fourier transformed (FFT) of the $\Delta R_{xy}$–$B^{-1}$ curves in (**b**), which exhibit two typical oscillatory components with the frequencies $F_\alpha = 51$ T and $F_\beta = 78$ T. Temperature dependences of $F_\alpha$ and $F_\beta$ are shown in the inset of (**c**). Dash lines are the fitting curves, exhibiting an exponential decay with increasing temperature. (**d**) Field dependent magnetoresistance $[(R_H-R_0)/R_0]$, where $R_H$ and $R_0$ represent the $R_{xx}$ taken at the magnetic fields of $\mu_0 H$ and zero-field, respectively.

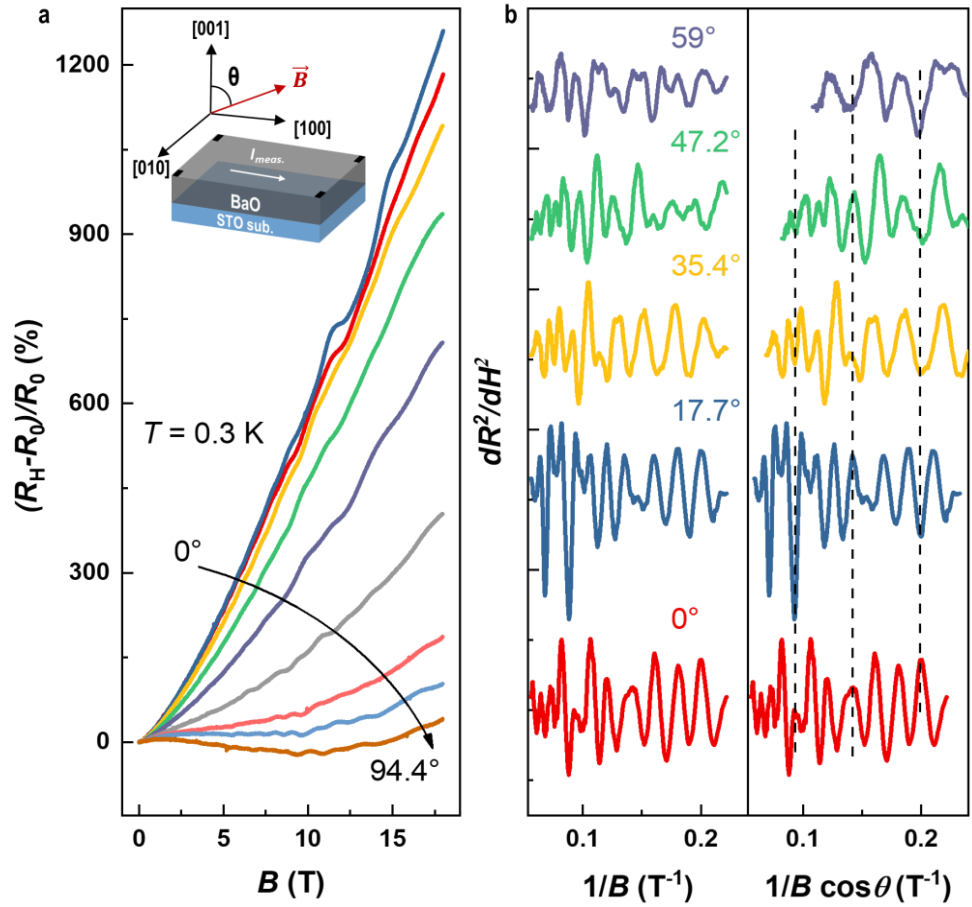

**Figure 4. Angular dependence of quantum oscillations.** (**a**) Magnetoresistance [($R_H$-$R_0$)/$R_0$] as a function of the magnetic fields at different angles ($\theta$) up to 18 T, where θ is the angle of applied field with respect to the surface normal. Inset shows the schematic of measuring setups. The current was applied in parallel to the [100] direction. All measurements were taken at 0.3 K. (**b**) Second derivative of resistivity ($dR^2/dH^2$), under different $\theta$ versus the reciprocal total magnetic field and the reciprocal perpendicular magnetic field component, respectively. The SdH oscillations depend on the reciprocal perpendicular magnetic field component, demonstrating the possible quasi-two-dimensional character of the Fermi surface present in the system.

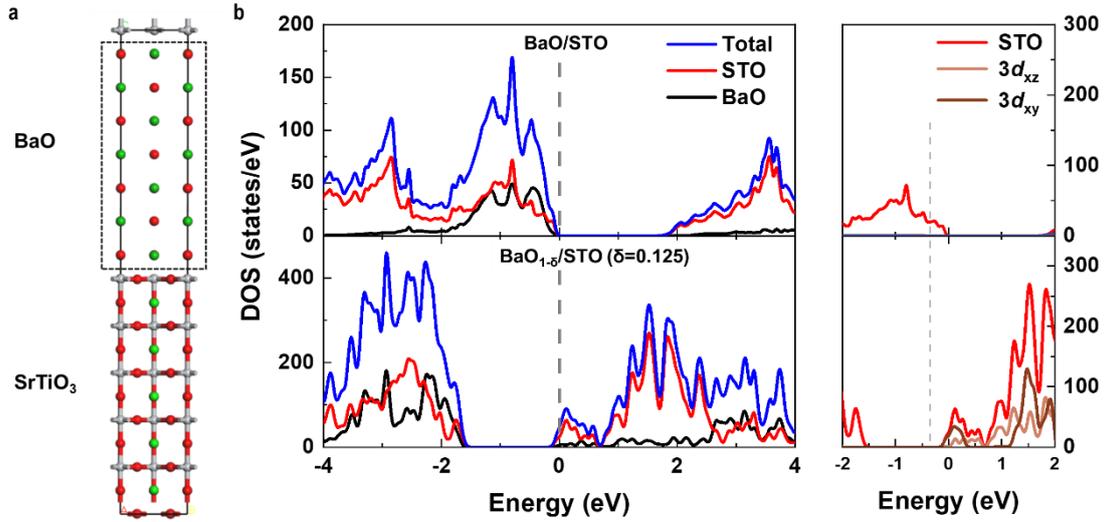

**Figure 5. Conducting interface between BaO and STO calculated using first-principles calculations.** (**a**) Differences in the charge density across the BaO/STO interface. The purple, grey, blue, and red balls represent the Ba, Sr, Ti, and O atoms, respectively. Yellow area corresponds to the hole accumulation, whereas the green area is for the electron accumulation. (**b**) Total and local Density of states (DOS) of BaO/STO interface without (upper panel) and with oxygen vacancies (lower panel) in BaO layer of the interface. The models are shown in Fig. S5 (**c**) The corresponding orbital resolved DOS for BaO/STO interface.